\documentclass[fleqn,usenatbib]{mnras}

\usepackage{newtxtext,newtxmath}
\usepackage[T1]{fontenc}
\DeclareRobustCommand{\VAN}[3]{#2}
\let\VANthebibliography\thebibliography
\def\thebibliography{\DeclareRobustCommand{\VAN}[3]{##3}\VANthebibliography}

\usepackage{color}
\usepackage{graphicx}
\usepackage{amsmath}
\usepackage{bm}
\usepackage[normalem]{ulem}

\title[Mind The Gap: High-$\beta$ Turbulent Cascades]{Mind The Gap: Nonlocal Cascades and Preferential Heating in High-$\beta$ Alfvénic Turbulence}

\author[W. Gorman and K. Klein]{
Waverly Gorman,$^{1}$\thanks{E-mail: waverlyg@arizona.edu}
Kristopher G. Klein,$^{1}$
\\
$^{1}$University of Arizona, Tucson, AZ, USA\\
}

\date{Accepted XXX. Received YYY; in original form ZZZ}

\pubyear{2023}

\begin{document}
\label{firstpage}
\pagerange{\pageref{firstpage}--\pageref{lastpage}}
\maketitle

\begin{abstract}
Characterizing the thermodynamics of turbulent plasmas is key to decoding observable signatures from astrophysical systems. In magnetohydrodynamic (MHD) turbulence, nonlinear interactions between counter-propagating Alfvén waves cascade energy to smaller spatial scales where dissipation heats the protons and electrons. When the thermal pressure far exceeds the magnetic pressure, linear theory predicts a spectral gap at perpendicular scales near the proton gyroradius where Alfvén waves become non-propagating. For simple models of an MHD turbulent cascade that assume only local nonlinear interactions, the cascade halts at this gap, preventing energy from reaching smaller scales where electron dissipation dominates, leading to an overestimate of the proton heating rate. In this work, we demonstrate that nonlocal contributions to the cascade, specifically large scale shearing and small scale diffusion, can bridge the non-propagating gap, allowing the cascade to continue to smaller scales. We provide an updated functional form for the proton-to-electron heating ratio accounting for this nonlocal energy transfer by evaluating a nonlocal weakened cascade model over a range of temperature and pressure ratios. In plasmas where the thermal pressure dominates the magnetic pressure, we observe that the proton heating is moderated compared to the significant enhancement predicted by local models.    
\end{abstract}




\section{Introduction and Background}
\label{sec:intro}
The presence of kinetic-scale plasma processes can be indirectly inferred in astrophysical systems through their influence on measurable quantities. One such example is the indirect detection of black holes through electron radiation from the turbulent plasma in the surrounding accretion disks \citep{EHT:2019,EHT:2022}. Understanding electron heating is critical to calibrate the expected radiation output and allow for the comparison of models to measurements \citep{Ressler:2015}. 

In general relativistic magnetohydrodynamic (GRMHD) simulations of accretion disks, the total proton-to-electron heating ratio $Q_p/Q_e$ directly affects the energy accessible to electrons. The energy deposited on the electrons is then distributed between various thermodynamic and radiative processes evolved in the simulations. \cite{Xie:2012} showed that the fraction of dissipation that heats the electrons strongly influences the radiative efficiency for a hot accretion flow. As an example, a comparison of $Q_p/Q_e$ prescriptions from the \cite{Howes:2011b} model for a Landau damped turbulent cascade and a model for heating from magnetic reconnection by \cite{Werner:2018} showed significant qualitative differences in properties of accretion flow as well as in simulated spectra and images between the two prescriptions when applied to simulations of Sgr A* \citep{Chael:2018} and jets around M87 \citep{Chael:2019}. Relative heating rates can be extracted from other heating mechanisms such as ion-cyclotron heating \citep{Squire:2023,Cranmer:2012} and stochastic heating \citep{Chandran:2010,Hoppock:2018,Cerri:2021} although $Q_p/Q_e$ prescriptions are not readily available. 
{A review of the parametric dependencies of these mechanisms can be found in \cite{Howes:2024}}. The current approach is to compare relative heating prescriptions with known dissipation mechanisms against observations to probe for the underlying physical mechanisms; see for example \cite{Dexter:2020} and \cite{Yao:2021}.  

The motivation for the present work is to modify the $Q_p/Q_e$ prescription for a critically-balanced Landau-damped turbulent cascade, increasing its application to a wider variety of astrophysical plasmas, specifically for systems where the proton thermal pressure dominates the magnetic pressure ($\beta_p>>1$) with $\beta_p \equiv \frac{2\mu_0n_pT_p}{B^2}$ (with $\mu_0$ the vacuum permeability, $n_p$ the proton number density, $T_p$ the proton temperature, and $B$ the magnetic field amplitude). The cascade model \citep[described in][]{Howes:2008b} follows the magnetic energy in a Landau-damped turbulent cascade as it transitions from magnetohydrodynamic (MHD) Alfvén waves to wavevector anisotropic low-frequency kinetic Alfvén waves, spanning inertial and dissipation ranges. The previous $Q_p/Q_e$ prescription \citep{Howes:2010} is a parametric study and fit of the \cite{Howes:2008b} model for varying $\beta_p$ and proton-to-electron temperature ratios ($\tau \equiv T_p/T_e$).
\begin{figure}
    \centering
    \includegraphics[width=0.9\linewidth]{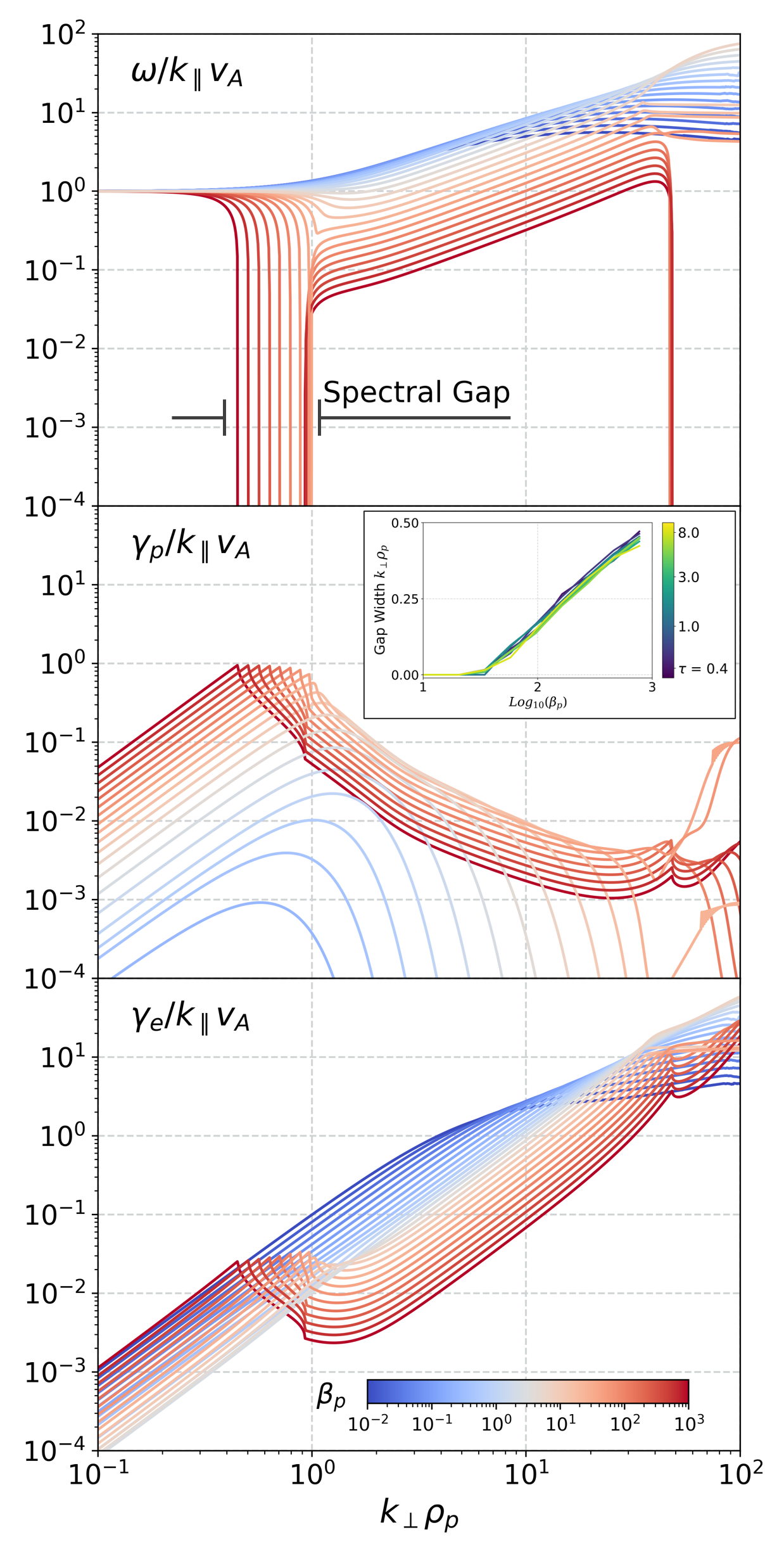}
    \caption{Real frequency $\omega/k_\parallel v_A$ (top panel) and damping rates $\gamma_s/k_\parallel v_A$ (proton - middle and electron - bottom) of the gyrokinetic Alfvén dispersion relation \citep{Howes:2006} for varying $\beta_p$ (color scale) with $\tau=1.0$. The spectral gap width as a function of $\log_{10}(\beta_p)$ is shown in the inset, with minimal variation as a function of $\tau$.}
    \label{fig:core4}
\end{figure}
Solutions to the linear dispersion relation of a collisionless gyrokinetic plasma in the high-$\beta_p$ limit ($\beta_p>>1$) develop a finite spectral gap where the real frequency $\omega$ approaches zero and the damping rate $\gamma$ reaches its local maximum around the proton gyroscale, $k_\perp\rho_p\sim$ 1 where $\rho_p$ is the proton Larmor radius\citep{Howes:2006, Hoppock:2018}. The gyrokinetic dispersion relation is a low-frequency anisotropic limit of kinetic theory \citep{Frieman:1982,Howes:2006} that depends on the perpendicular spatial scale and plasma conditions. The total damping rate can be decomposed into contributions from each particle species, $\gamma(k_\perp, \beta_p,\tau)=\sum_{s}\gamma_s(k_\perp, \beta_p, \tau)$. The middle and bottom panels of Fig.~\ref{fig:core4} show the proton and electron $\gamma_s$ as a function of perpendicular wavevector ($k_\perp$) and $\beta_p$ for fixed $\tau=1.0$. While the electron damping is only minimally affected by $\beta_p$, the proton damping grows with $\beta_p$, reaching a local maximum at the spectral gap near the proton gyroscale. As shown in the top panel of Fig.~\ref{fig:core4}, the gap begins to form around $\beta_p$=30 and increases in width for increasing $\beta_p$. We found the gap width is largely independent of $\tau$ for the range included in this study, as shown in the Fig.~\ref{fig:core4} inset. 

A local cascade, which relies on interactions between counter-propagating Alfvén waves of similar size to transfer energy across scales, breaks down when energy reaches the spectral gap. The formation of the gap is described in \cite{Kawazura:2018} and briefly summarized here. Alfvén waves are damped at a rate that peaks around $k_\perp\rho_p \sim \beta_p^{-1/4}$ for large values of $\beta_p$. This damping is sufficient to slow, and eventually stop at high enough values of $\beta_p$, the propagation of Alfvén waves with scales around the proton gyroradius, $k_\perp\rho_p\sim$ 1. Past the proton scales the damping diminishes, re-emerging at much smaller scales when electron Landau damping begins to dominate. \cite{Howes:2011c} developed an update to the local cascade model with additional channels of energy transfer via large-scale shearing and small-scale diffusion, but the $Q_p/Q_e$ prescription varying $\beta_p$ and $\tau$ \citep{Howes:2010} has not been updated to include these processes. New fits of $Q_p/Q_e$ with this physics included are provided in this work, Section~\ref{sec:results}. We show that including these nonlocal energy transfer mechanisms in the updated model is sufficient to bridge the gap and allow the cascade to continue to smaller scales, enabling higher rates of relative electron heating. The nonlocal energy transfer mechanisms are described in \citep{Howes:2011c} and briefly reviewed in Section~\ref{ssec:cascade}. The three models compared in this work are the "original" Local model, the updated Weakened Nonlocal model and the Weakened Local model, where the nonlocal energy transfer is switched off but the scale dependent variation between the linear and nonlinear timescales is preserved.

This paper is organized as follows: Section~\ref{sec:method} describes the three cascade models, the inputs used for this analysis, and the process for fitting $Q_p/Q_e$ from each of the models including the evaluation of the goodness of fit. Section~\ref{sec:results} provides the best fitting parameters for each model and discusses the results. Findings and key takeaways are summarized in Section~\ref{sec:conclusions}.

\section{Methodology}
\label{sec:method}

\subsection{Cascade Models}
\label{ssec:cascade}

This work compares three models: Local \citep{Howes:2008b}, Weakened Local, and Weakened Nonlocal \citep{Howes:2011c}. These models describe a steady-state cascade of magnetic energy from inertial through dissipation scales where Landau damping terminates the cascade onto protons and electrons.

The Local model uses a Batchelor-like\citep{Batchelor:1953}, one-dimensional continuity equation for the magnetic energy spectrum $b_k^2$ in perpendicular wave number $k_\perp$ space,
\begin{equation}
    \frac{\partial b_k^2}{\partial t}=-k_\perp\frac{\partial\epsilon(k_\perp)}{\partial k_\perp}+s(k_\perp)-2\gamma b_k^2
    \label{eqn:Batchelor}
\end{equation} 
that assumes only local nonlinear energy transfer; i.e. interactions occur only between fluctuations of similar size. It also assumes critical balance between the linear propagation and nonlinear interaction times; i.e. the ratio of timescales is maintained as the cascade proceeds to smaller scales\citep{GoldreichSridhar:1995, Schekochihin:2022}. The three contributions to this spectrum are the flux of energy in wave number space (the first term on the right-hand side of Eqn.~\ref{eqn:Batchelor}), the source term for driving the turbulence $s(k_\perp)$ which injects energy at the largest scales, and a damping term that scales with the linear damping rate $\gamma$. By assuming steady state, Eqn.~\ref{eqn:Batchelor} can be integrated to find the energy cascade rate $\epsilon$ as a function of $k_\perp$, $\beta_p$, and $\tau$. The scale-dependent species heating rate can be expressed as a function of the cascade and damping rates, 
\begin{equation}
    Q_{k,s}=2 K_1^{3/2} K_2 \frac{\gamma_s}{\omega}\frac{\epsilon(k_\perp)}{k_\perp}
\end{equation}
where $K_j$ are dimensionless Kolmogorov constants; see \cite{Howes:2010} for additional details.

\begin{figure*}
    \centering
    \includegraphics[width=\linewidth]{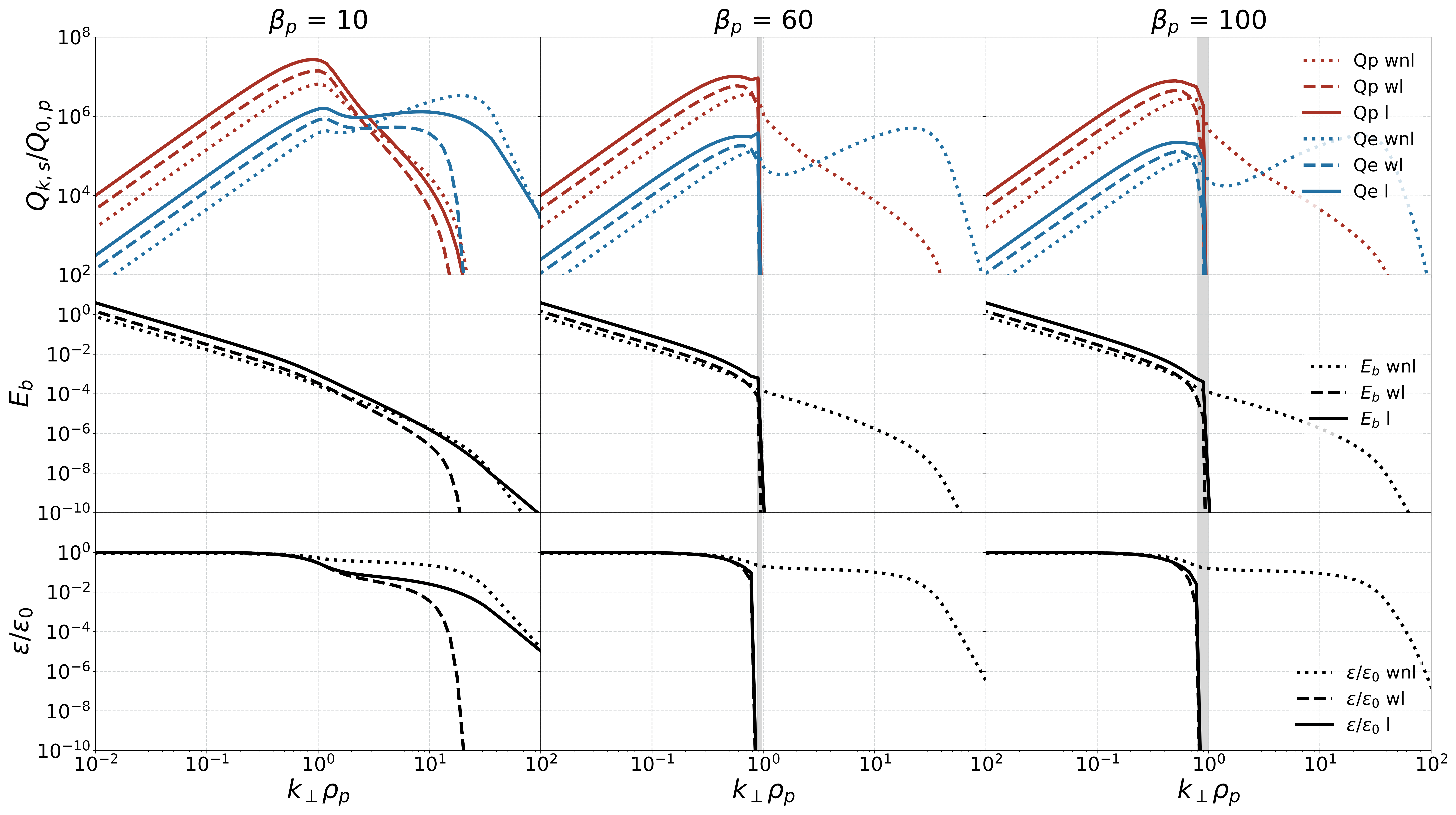}
    \caption{Heating rates $Q_{k,s}$ normalized by the outer scale proton rate $Q_{0,p}$ (top row), energy spectra $E_b$ (center row) and cascade rates $\epsilon$ normalized by the outer scale cascade rate $\epsilon_0$(bottom row) from the Local (solid), Weakened Local (dashed) and Weakened Nonlocal (dotted) cascade models as a function of $k_\perp$ for three values of $\beta_p$; low $\beta_p$=10 (left column), moderate $\beta_p$=60 (center column), and high $\beta_p$=100 (right column) for $\tau=1.0$. The gray bars illustrate the $\omega$=0 region---the spectral gap--- for the $\beta_p=60$ and $100$ cases. The presence of this gap leads to a termination of the local cascade and an inflation of the relative proton heating rates.}
    \label{fig:cube}
\end{figure*}

The original cascade model \citep{Howes:2008b} predicted an exponential fall-off in the spectrum due to strong linear kinetic damping at sub-ion-Larmor scales. This fall-off has not been observed in most observations of turbulence in the solar wind (e.g., \cite{Chen:2010,Sahraoui:2010}), thus, a weakened cascade model was proposed by \cite{Howes:2011c} as a refinement to the original Local model to correct for this discrepancy. The term weakened comes from relaxing the assumption of critical balance as the cascade progresses: the relative linear and nonlinear timescales are allowed to vary as a function of scale. The Weakened Local model maintains the same assumption of local energy transfer as in the original cascade model. 

The Weakened Nonlocal model allows for nonlocal contributions to the nonlinear energy transfer by providing a quantification for shearing (where large eddies shear apart much smaller eddies) and diffusion (where small eddies diffuse across much larger eddies). Nonlocal contributions enable the cascade to bridge the gap region when the Alfv\'en frequency is zero, and thus the local cascade rate is also zero. As these contributions persist at all scales, we do expect to find small deviations in $Q_p/Q_e$ between the Local and Nonlocal models even for $\beta_p<30$. 

Fig.~\ref{fig:cube} shows a comparison between the three models for low, moderate, and high $\beta_p$ cases. We classify $\beta_p$=10 as "low" as it is below the $\beta_p$=30 threshold for forming the zero-frequency gap. For the moderate and high $\beta_p$ cases, a finite gap width results in the cascade halting near $k_\perp \rho_p$ for both local models (shown in the bottom row).
This interrupted cascade reshapes the steady state magnetic energy spectrum $E_b$, (middle row) truncating it at proton kinetic scales.
This truncation modifies the scale-dependent particle heating (top row; proton - red, electron - blue) . 
All heating for the high-$\beta_p$ local models stops at the onset of the gap, where proton heating is maximum. The nonlocal transfer of energy to smaller scales enables an enhancement of the electron heating. 

The total proton-to-electron heating rate, $Q_p/Q_e$, is extracted by integrating $Q_{k,s}$ over $k_\perp$ to yield $Q_s$ and taking the ratio of the energy that goes into protons by the energy that goes into electrons. Thus, although $Q_{k,e}$ approaches zero at the gap for the high $\beta_p$ cases, the integral $Q_e$ approaches a steady value allowing the ratio to asymptote to a finite value. The ratio $Q_p/Q_e$ is calculated over a range of $\beta_p$ and $\tau$ for each model. A finite gap width halts the Local cascade model around $k_\perp\rho_p \sim 1$, thus no energy reaches electron scales, leading to a significant enhancement of $Q_p/Q_e$.

\subsection{Simulation Details}
\label{methodtool}
The three cascade models are calculated using a \texttt{FORTRAN} code developed by \cite{Howes:2008b} which has been updated to include the Weakened Nonlocal and Weakened Local models\citep{Howes:2011c}. Input parameters used for this effort include the proton to electron mass ratio, fixed to 1836; the Kolmogorov constants $C_1$, and $C_2$, fixed to 1.9632 and 1.0906 respectively; and the initial and final $k_\perp$ sweep parameters, fixed to 0.0001 and 150 respectively. The resolution of $k_\perp$ is set to 100 points and was reduced if needed to ensure numerical stability of the code. A convergence study was conducted to ensure that slight variation in the number of points in $k_\perp$ did not have a significant impact on the resulting $Q_p/Q_e$. The range for $k_\perp$ was selected to allow for coverage of all scales of interest, spanning scales larger than the proton gyroscale (around the region where the gap occurs) and well into electron scales. The choice of Kolmogorov constants influence the results of cascade calculation; see \cite{Howes:2008b} for the impacts on the predicted spectrum and \cite{Shankarappa:2023} where the constants were fit to measured data from Parker Solar Probe. We choose our values to align with \cite{Howes:2011c}. The Weakened cascade model defines a third Kolmogorov constant $C_3$ which allows for normalization of the nonlinear frequency for the Weakened Local model of the form
\begin{equation}
\omega_{nl}(k_\perp)=C_3\omega_{nl}^{(loc)}(k_\perp).
    \label{eqn:kc3defn}
\end{equation}
We use the same Kolmogorov constants as \cite{Howes:2011c}, setting $C_3 = 2.25$ for the Weakened Local model to allow direct comparison to the Weakened Nonlocal results. 

\begin{figure*}
    \centering
    \includegraphics[width=0.8\linewidth]{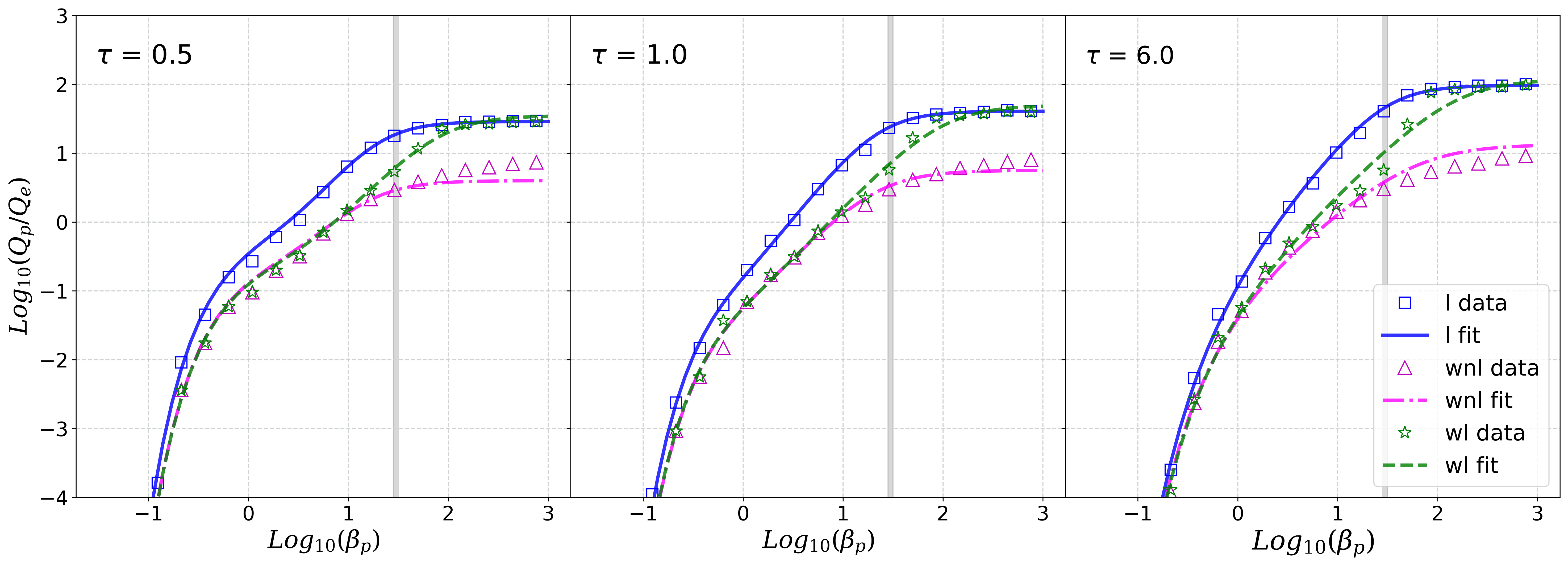}
    \caption{The total proton-to-electron heating ratio $Q_p/Q_e$ as a function of $\log_{10}(\beta_p)$ for three values of $\tau$: 0.5 (left), 1.0 (center), and 6 (right) for the three cascade models; Local (blue square), Weakened Local (green star) and Weakened Nonlocal (pink triangle). The model data are evaluated as the binned average values, described in Section~\ref{ssec:mcmc}, and the fit is evaluated using the best fit parameters from Table~\ref{table:modcompare}. Above the gap threshold of $30$ (grey vertical bar), the Weakened Nonlocal model has a significantly reduced $Q_p/Q_e$ while the Weakened Local model converges to the Local model.}
    \label{fig:slices}
\end{figure*}

We consider values of $\beta_p$ between 0.02 and 1000 and $\tau$ between 0.2 and 10. The values for $\beta_p$ and $\tau$ where chosen to be consistent with previous work, with an extension to higher $\beta_p$ values to better explore parameters with larger gap widths. The lower values of $\beta_p$ were selected such that $Q_p/Q_e$ is effectively zero to machine precision. The lower range for $\tau$ was limited by impacts of couplings between the Alfv\'en and compressive modes at extreme temperature ratios.
Examples of the calculated $Q_p/Q_e(\beta_p)$ for the three cascade models for fixed $\tau$ values are shown in Fig.~\ref{fig:slices}.
As expected from the analysis of Fig.~\ref{fig:cube}, $Q_p/Q_e$ is significantly reduced for the Nonlocal model at high values of $\beta_p$ above the gap threshold of $30$.

\subsection{Fitting and Error Estimates}
\label{ssec:mcmc}
We computed $Q_p/Q_e$ at 1300 points randomly distributed in $\log_{10}(\beta_p)$ and $\log_{10}(\tau)$ space.
A binned histogram of the relative heating rates is shown in Fig.~\ref{fig:ModData} to visualize the data. The raw data comprised of the 1300 $(\beta_p,\tau)$ pairs were used for fitting.
We chose to use a random set of points instead of a fixed grid as we found that the random set had slightly faster convergence of the fit with minimal effect on the resulting prescription. 

The $Q_p/Q_e$ data for each model is fit using least-squares from the LMFIT python module by \cite{Newville:2014} and 
the functional form used by \cite{Howes:2011b} 
\begin{equation}
\frac{Q_p}{Q_e}=c_1\frac{(c_2/\tau)^2+\beta_p^{(\alpha_1-\alpha_2)}}{(c_3+c_4log(\tau))^2+\beta_p^{(\alpha_1-\alpha_2log(\tau))}}\sqrt{\frac{m_p}{m_e}\tau}\:e^{-1/\beta_p}.
    \label{eqn:howes_p}
\end{equation}
This function has six free parameters: $c_1, c_2$, $c_3, c_4, \alpha_1, \alpha_2$. 

The LMFIT module provides a reduced $\chi^2$ for each fit, calculated as the sum of the squared residuals divided by the number of points (1300), which is included in Table~\ref{table:modcompare} in Section~\ref{sec:results}.  

\section{Results and Discussion}
\label{sec:results}
\begin{figure}
    \centering
    \includegraphics[width=0.8\linewidth]{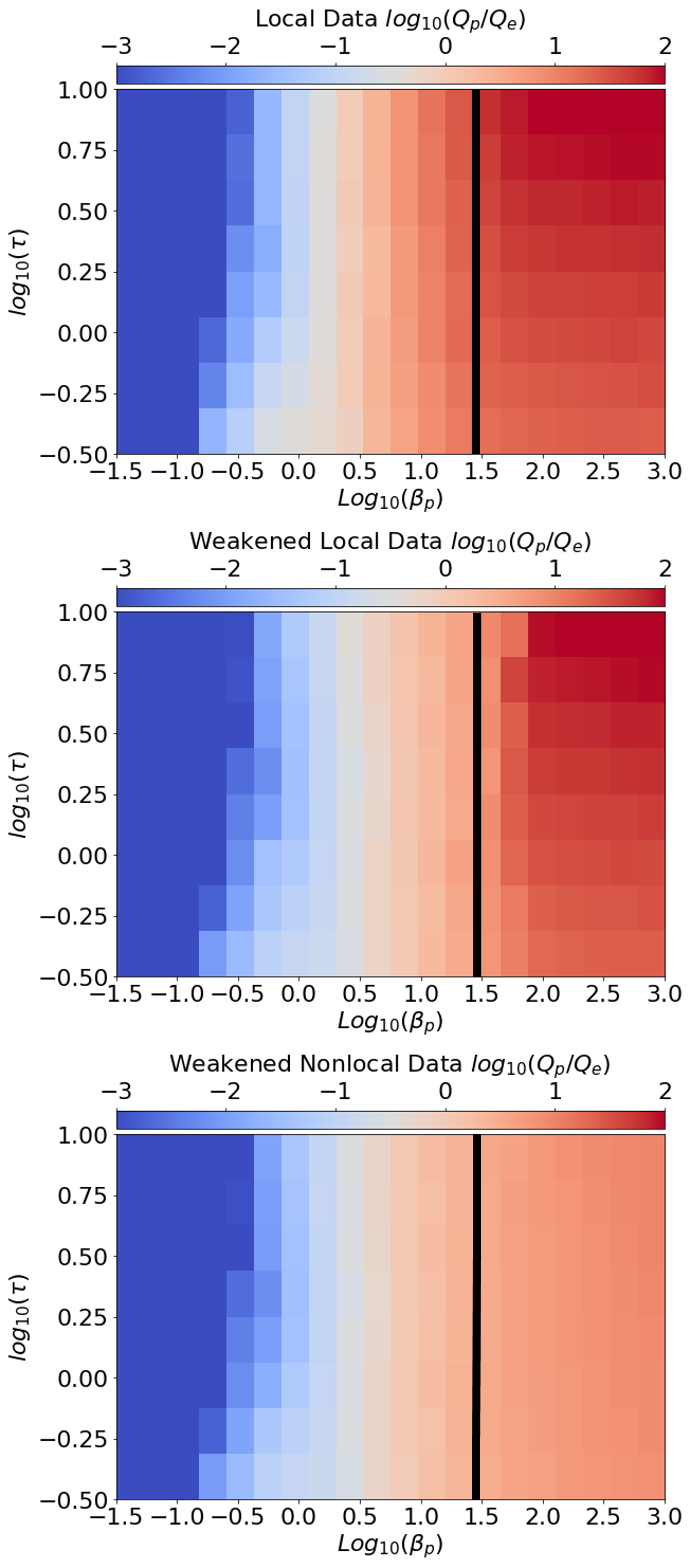}
    \caption{Local (top), Weakened Local (middle), and Weakened Nonlocal (bottom) relative heating rates $Q_p/Q_e$ binned as a function of $\beta_p$ and $\tau$. The onset of the gap is shown with a black line. The relative heating below the gap is similar for the two Weakened models while above the gap the two local models are similar. At high $\beta_p$, the nonlocal contributions to the cascade result in a significant suppression in $Q_p/Q_e$.}
    \label{fig:ModData}
\end{figure}

\begin{table}
\centering
\caption{Best fit parameters in the form $[c_1,c_2,c_3,c_4,\alpha_1,\alpha_2]$ for Eqn.~\ref{eqn:howes_p} with the reduced $\chi^{2}$ for the new fit; the reduced $\chi^2$ using the original local model prescription parameters from \citet{Howes:2010} $[0.92, 1.6, 18, 5, 2, 0.2]$ is given in parenthesis.}
\begin{tabular}{cc}
  \textbf{Model} & \textbf{New Best Fit} \\
  \hline
  Local & [0.926, 1.21, 14.8, 3.90, 1.73, 0.052] \\
   & $\chi^2=0.0040$ (0.0088) \\
  Weakened Local & [1.128, 1.371, 29.38, 4.363, 1.477, 0.124] \\
   & $\chi^2=0.0160$ (0.2261) \\
  Weakened Nonlocal & [0.128, 1.267, 9.40, 2.61, 1.47, 0.368] \\
   & $\chi^2=0.0149$ (0.5305) \\
  \hline
\end{tabular}
  \label{table:modcompare}
\end{table}

The results of the fitting procedure described in Section ~\ref{sec:method} are summarized in Table~\ref{table:modcompare}. The reduced $\chi^2$ is provided for each fit.

The smaller $\chi^2$ values show the new fits are slightly better than the previous fit for the local models. One possible explanation for the improvement could be that the present fit was performed with 6 free parameters while the original fit was performed with less than 6 (the original fit combined $c_3$ with $c_4$ and $\alpha_1$ with $\alpha_2$). The present fit was performed with 1300 randomly spaced points in $\beta_p$, $\tau$ space and it is unknown how many points were used in the original fit. Additionally, we used the LMFIT implementation of the emcee package \citep{MCMC:2013} to optimize the initial guess parameters for the fit and make sure that each parameter was well sampled.

Model relative heating rates as a function of $\beta_p$ and $\tau$ are shown in Fig.~\ref{fig:ModData}. 
Nonlocal energy transfer mechanisms, which enabled energy transfer across the gap, moderated proton heating and reduced the $Q_p/Q_e$ compared to local models. Thus, we see the suppression of $Q_p/Q_e$ for the Weakened Nonlocal model beginning at the gap threshold of $\beta_p~=30$. 

\subsection{Impact of Kolmogorov Constants}
\label{ssec:constants}

\begin{figure*}
    \centering
    \includegraphics[width=0.95\linewidth]{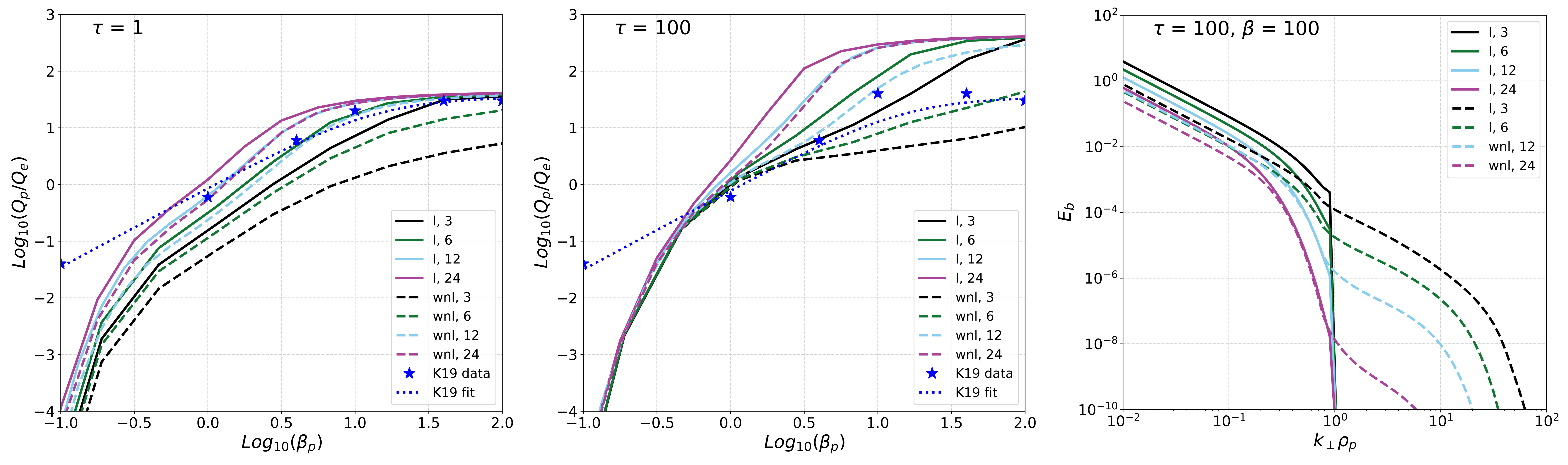}
    \caption{The total proton-to-electron heating ratio $\log_{10}(Q_p/Q_e)$ as a function of $\log_{10}(\beta_p)$ for two values of $\tau$: 1 (left) and 100 (center) varying the product of Kolmogorov constants $C_1^{3/2}C_2$ = 3, 6, 12, and 24 compared to fully nonlinear gyrokinetic simulations conducted by \protect\cite{Kawazura:2019} (K19). For the fits described in Section ~\ref{sec:method}, the value of $C_1^{3/2}C_2$ was fixed to 3 (black lines). Increasing this product results in a steepening of the magnetic energy spectra (right, for $\tau=100$ and $\beta_p=100$) effectively reducing the impact of the spectral gap on the heating ratio at high-$\beta_p$.}
    \label{fig:k18compare}
\end{figure*}

We held the product of the Kolmogorov constants fixed to $C_1^{3/2}C_2=3$ for consistency with \cite{Howes:2008b, Howes:2010, Howes:2011b}. This is reasonable under the assumption that these constants do not vary significantly over the range of $\beta$ and $\tau$ covered in this work. To investigate this assumption, we compared two cases, $\tau$ = 1 and $\tau$ = 100 for the Local and Weakened Nonlocal models to fully nonlinear gyrokinetic simulations by \cite{Kawazura:2019} (referred to as K19) shown in Fig.~\ref{fig:k18compare}. The K19 results are directly extracted from simulations, and thus do not assume a particular value of $C_1^{3/2}C_2$. The simulations in K19 covered a large range of $\beta_p$ and $\tau$ and observed that $Q_p/Q_e$ asymptoted to a constant $\simeq$30 for high-$\beta_p$ systems. Figures~\ref{fig:slices} and ~\ref{fig:k18compare} show that while all three cascade models converge to a limiting value of $Q_p/Q_e$ that increases slightly for increasing values of $\tau$, the Weakened Nonlocal model converges to a much lower value, e.g. $\sim$8 for $\tau = 1.0$. We found that increasing $C_1^{3/2}C_2$ led to better agreement with the K19 results for the $\tau=1$ case, but there was not a consistent value that matched well across $\beta_p$ for the $\tau=100$ case. Comparing the Weakened Nonlocal models for $\beta_p > 1$, $C_1^{3/2}C_2$ = 6 matched well for the $\tau$ = 100 case while the $C_1^{3/2}C_2$ = 12 matched well for $\tau$ = 1. 

Increasing the product of Kolmogorov constants, shown to increase the limiting value of $Q_p/Q_e$, effects the turbulent cascade by slowing down the rate at which energy is transferred from scale to scale. This has the net effect of providing more time for linear damping to occur and remove energy from the cascade before it reaches smaller scales, thus reducing the effect of the spectral gap, which halts any further energy transfer past the proton gyroscale, if the product is sufficiently large. This can be seen in Fig.~\ref{fig:k18compare} where large products have a much smaller difference between the Local and Weakened Nonlocal models and on the right panel where the Weakened Nonlocal magnetic energy spectra appears to approach the Local value as the product increases. As we showed, the heating ratio $Q_p/Q_e$ has a significant dependence on the choice of Kolmogorov constants, thus, any application of this class of prescription must consider if the underlying Kolmogorov constants are appropriate for the plasma system of interest. Notably, a recent study attempting to observationally constrain lower-$\beta_p$ intervals using in situ magnetic field spectral measurements found significant variability in these constants (c.f. Fig. 2 from \cite{Shankarappa:2023}). By separately extracting the best fit $C_1$ and $C_2$ for the first two Parker Solar Probe encounters, \cite{Shankarappa:2023} found a median $C_1^{3/2}C_2$ of 45.7 (1\textsuperscript{st} quartile = 29.6, 3\textsuperscript{rd} quartile= 74.0); and median of 39.7 (1\textsuperscript{st} quartile = 26.2, 3\textsuperscript{rd} quartile = 56.9), for Encounters 1 and 2 respectively. 

\subsection{Impact of Pressure Anisotropy}
\label{ssec:anisotropy}

\begin{figure*}
    \centering
    \includegraphics[width=\linewidth, trim=1cm 1.05cm 0.cm 0.25cm, clip]{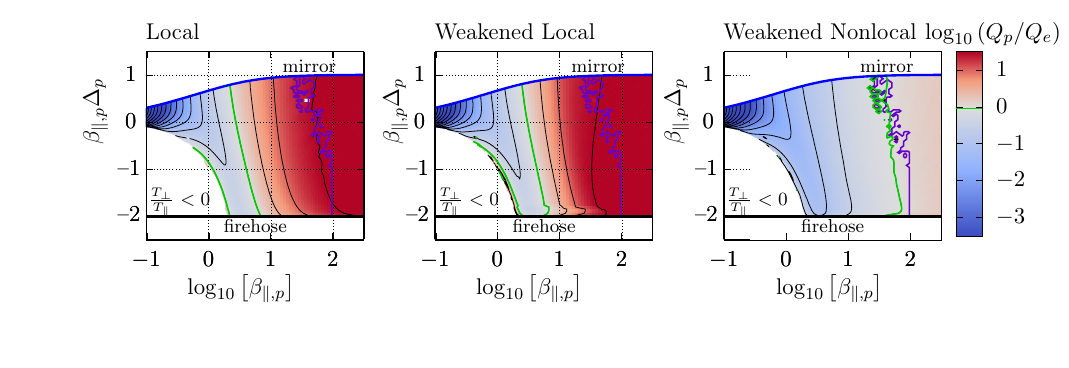}
    \caption{The total proton-to-electron heating ratio $Q_p/Q_e$ as a function of $\beta_{\parallel,p}$ and weighted proton pressure anisotropy for the Local, Weakened, and Nonlocal models with $\tau=1$ and $\Delta_e=0$. The same format is used as Fig. 6 from \protect\cite{Kunz:2018}, with an extended range of $\beta_{\parallel,p}$. The green line indicates $Q_p = Q_e$, with the purple line illustrating the numerically identified location of the spectral gap. For all three models, $Q_p/Q_e$ is only weakly dependent on $\Delta_p$, especially in the thermally dominated regime.}
    \label{fig:anisotropy}
\end{figure*}

High-$\beta$ plasmas are particularly susceptible to the occurrence of pressure anisotropies \citep{Schekochihin:2008,Foucart:2017,Kempski:2019}, which can be usefully quantified by $\Delta_s\equiv T_{\perp,s}/T_{\parallel,s}-1$.
Pressure anisotropies can impact the transport and dynamics of such systems, as demonstrated in a variety of numerical simulations \citep{Riquelme:2015,Kunz:2016,Arzamasskiy:2023}.
To characterize the impact of such anisotropies on the nonlocal cascade calculation presented in this work, we repeat the calculations laid out in \cite{Kunz:2018}, where the modifications to the gyrokinetic dispersion relation and associated damping rates are derived and applied to the local cascade model, with the Weakened Local and Nonlocal cascade models. 
The results as a function of $\beta_{\parallel,p}$ and $\beta_{\parallel,p}\Delta_p$ are illustrated in Fig.~\ref{fig:anisotropy}.
We vary the proton pressure anisotropy between the CGL firehose and mirror thresholds and cover an extended range of high-$\beta_{\parallel,p}$ values.
The electron pressure anisotropy is set to zero $\Delta_e=0$, and $\tau$ is fixed to unity.
Especially for high-$\beta_{\parallel,p}$ region, the relative heating rate is only weakly controlled by $\Delta_p$. 
For these cascade models, as long as the pressure anisotropy is within the bounds of stability, the other system parameters have a more significant impact on the bifurcation of energy between the plasma components.

\subsection{Prescription Limitations}
\label{ssec:limits}

In Section~\ref{ssec:constants} we demonstrated the impact of the choice of Kolmogorov constants on the derived heating rates. There are a number of additional assumptions built in to these cascade models that influence how the prescription should be applied that may need modification depending on the plasma system of interest. These models assume a single isotropic driving scale $k_{\perp,0}$, use linear gyrokinetic damping rates $\gamma$ to describe the process of removing energy from the cascade, and ignore power in compressive modes \citep{Kawazura:2020}. These models only describe the rate of energy removal associated with linear Landau damping. We neglect to address how the energy is processed once it is removed from the turbulent fluctuations and transferred to the particle distribution functions, placing questions about phase mixing and collisional heating between particles \citep{Kanekar:2015,Schekochihin:2016} outside the scope of this work. Extracting the linear damping rates derived using the gyrokinetic ordering neglects other channels of energy dissipation including cyclotron heating, stochastic heating, or magnetic reconnection as well as the impact of quasilinear deformations to the velocity distribution \citep[e.g.][]{Isenberg:2012}. A discussion of how these various assumptions affect the cascade model used in this work as applied to the slow solar wind can be found in \cite{Howes:2008b}. 

\section{Conclusion}
\label{sec:conclusions}
We compared models for a Landau-damped turbulent cascade with and without nonlocal contributions to the energy transfer. Nonlocal interactions as implemented in the Weakened Cascade Model by \cite{Howes:2011c} are sufficient to bridge the gap in wavenumber space for high-$\beta$ plasmas where Alfvén waves become non-propagating, allowing the cascade to continue to electron scales. This paper provides updated fits to the functional form for $Q_p/Q_e$ proposed by \cite{Howes:2010} allowing nonlocal contributions. The functional form of \cite{Howes:2010} appears to work well for the Weakened Nonlocal model. Finally, we showed that in plasmas where the thermal pressure dominates the magnetic pressure, local models overpredict the relative proton heating; this effect is caused by the halting of the cascade near the proton gyroscale due to the zero-frequency gap resulting in a larger $Q_p/Q_e$ at electron scales.

\section{Data Availability}
\label{sec:data}
The data that support the findings of this study are available from the corresponding author, W.G., upon reasonable request.

\section{Acknowledgements}
\label{sec:thanks}
K.G.K. is supported by NASA Grants 80NSSC19K0912 and 80NSSC20K0521. 
The authors thank G. Howes for providing the \texttt{FORTRAN} code developed for the calculation of the cascade models described in \cite{Howes:2008b} and \cite{Howes:2011c}. 
The authors also thank the referee for insightful comments that have improved the manuscript.

\bibliographystyle{mnras}
\bibliography{main}

\bsp	
\label{lastpage}
\end{document}